\documentclass[a4paper,fleqn,usenatbib]{mnras}

\usepackage{newtxtext,newtxmath}
\usepackage[T1]{fontenc}
\usepackage{ae,aecompl}

\usepackage{graphicx}	% Including figure files
\usepackage{amsmath}	% Advanced maths commands  %To put text mode inside math mode
\usepackage{amssymb}	% Extra maths symbols

\usepackage[flushleft]{threeparttable}
\usepackage[pdftex,usenames,dvipsnames]{xcolor}
\usepackage{natbib}
\usepackage{aas_macros}
\usepackage{url}
\usepackage[none]{hyphenat}
\usepackage{times}
\title[Extended radio jet of ULX]{The extended radio jet of an off-nuclear low-mass AGN in NGC 5252}

\author[Mezcua et al.]{
M. Mezcua,$^{1,2}$\thanks{E-mail: marmezcua.astro@gmail.com}
M. Kim,$^{3,4,5}$
L.C. Ho$^{6,7}$
C.J. Lonsdale$^{8}$
\\
$^{1}$Institute of Space Sciences (ICE, CSIC), Campus UAB, Carrer de Magrans, 08193 Barcelona, Spain\\
$^{2}$Institut d'Estudis Espacials de Catalunya (IEEC), Carrer Gran Capit\`{a}, 08034 Barcelona, Spain\\
$^{3}$Korea Astronomy and Space Science Institute, Daejeon 305-348, Korea\\
$^{4}$University of Science and Technology, Daejeon 305-350, Korea\\
$^{5}$Department of Astronomy and Atmospheric Sciences, Kyungpook National University, Daegu  702-701, Korea\\
$^{6}$Kavli Institute for Astronomy and Astrophysics, Peking University, Beijing 100871, China\\
$^{7}$Department of Astronomy, School of Physics, Peking University, Beijing 100871, China\\
$^{8}$National Radio Astronomy Observatory, Charlottesville, VA 22903, USA
}

\date{Accepted XXX. Received YYY; in original form ZZZ}

\pubyear{2015}

\begin{document}
\label{firstpage}
\pagerange{\pageref{firstpage}--\pageref{lastpage}}
\maketitle

\begin{abstract}
CXO J133815.6+043255 is an ultraluminous X-ray source (ULX) with ultraviolet, optical, and radio counterparts located 10 kpc away from the nucleus of the galaxy NGC 5252. Optical spectroscopic studies indicate that the ULX is kinematically associated with NGC 5252; yet, the compactness of its radio emission could not rule out the possibility that the ULX is a background blazar. We present follow-up VLBA radio observations that are able to resolve the compact radio emission of the ULX into two components, making the blazar scenario very unlikely. The east component is extended at 4.4 GHz and its detection also at 7.6 GHz reveals a steep spectral index. The west component is only detected at 4.4 GHz, is not firmly resolved, and has a flatter spectral index. Considering that the west component hosts the radio core, we constrain the black hole mass of the ULX to $10^{3.5} < M_\mathrm{BH} \lesssim 2 \times 10^{6}$ M$_{\odot}$ and its Eddington ratio to $\sim 10^{-3}$. The ULX is thus most likely powered by an intermediate-mass black hole or low-mass AGN. Our results constitute the first discovery of a multi-component radio jet in a ULX and possible intermediate-mass black hole.
\end{abstract}


%%%%%%%%%%%%%%%%%%%%%%%%%%%%%%%%%%%%%%%%%%%%%%%%%%

%%%%%%%%%%%%%%%%% BODY OF PAPER %%%%%%%%%%%%%%%%%%

\section{Introduction}
Ultraluminous X-ray sources (ULXs) were initially touted as sub-Eddington accreting intermediate-mass black holes (IMBHs) with BH masses $100 \lesssim M_\mathrm{BH} \lesssim 10^{5}$ M$_{\odot}$ because of their location off the center of galaxies and their X-ray luminosities exceeding the Eddington limit of a 10 M$_\odot$ stellar-mass BH ($L_\mathrm{X} \geq 10^{39}$ erg s$^{-1}$; e.g., see review by \citealt{2017ARA&A..55..303K}). The recent finding of X-ray pulsations in some ULXs (\citealt{2014Natur.514..202B}; \citealt{2017Sci...355..817I,2017MNRAS.466L..48I}), together with dynamical mass measurements (\citealt{2013Natur.503..500L}) indicate that many of them are instead either stellar-mass BHs or neutron stars accreting at super-Eddington rates.

Only those extreme ULXs with $L_\mathrm{X} \geq 5 \times 10^{40}$ erg s$^{-1}$, not easily explained by super-Eddington accretion, remain as possible IMBH candidates (see review by \citealt{2017IJMPD..2630021M}). This is the case of HLX-1 (e.g., \citealt{2009Natur.460...73F}; \citealt{2011ApJ...734..111D}; \citealt{2012Sci...337..554W}), tagged as the best IMBH candidate among ULXs, M82-X1 (\citealt{2001MNRAS.321L..29K}; \citealt{2014Natur.513...74P}), or NGC 2276-3c (\citealt{2012MNRAS.423.1154S}; \citealt{2013MNRAS.436.3128M,2015MNRAS.448.1893M}), three ULXs suggested to be the nucleus of a dwarf galaxy stripped during a minor merger with the ULX host galaxy (\citealt{2005MNRAS.357..275K}; \citealt{2013ApJ...768L..22S}; \citealt{2015MNRAS.448.1893M}). This scenario adds to the growing body of evidence that IMBHs or low-mass AGN ($M_\mathrm{BH} \lesssim 10^{6}$ M$_{\odot}$) can be found in dwarf galaxies (e.g., \citealt{2003ApJ...588L..13F}; \citealt{2004ApJ...607...90B}; \citealt{2004ApJ...610..722G,2007ApJ...670...92G}; \citealt{2013ApJ...775..116R}; \citealt{2015ApJ...809L..14B,2017ApJ...836...20B}; \citealt{2017ApJ...836..237N}; \citealt{2016ApJ...817...20M,2018MNRAS.478.2576M}), which has strong implications for understanding how supermassive BHs form. The finding of high-redshift quasars when the Universe was only 0.7 Gyr (e.g., \citealt{2011Natur.474..616M}; \citealt{2015Natur.518..512W}; \citealt{2018Natur.553..473B}) and of ultramassive BHs of more than 10$^{10}$ M$_\odot$ in the local Universe (\citealt{2011Natur.480..215M}; \citealt{2018MNRAS.474.1342M}) suggests that these behemoths must have been seeded by BHs of 10$^{2}-10^{5}$ M$_\odot$ in the early Universe and then grow via accretion and cosmological merging (\citealt{2003ApJ...582..559V}). Theoretical models predict that the leftover of those seed BHs that did not grow into supermassive should be found in local dwarf galaxies (e.g., \citealt{2010MNRAS.408.1139V}), where they might shine as ULXs when undergoing a minor merger that strips the dwarf galaxy of its stellar body. CXO J133815.6+043255 is a recently discovered ULX that bolsters this possibility. 

The ULX CXO J133815.6+043255 is located at a projected separation of 22 arcsec ($\sim$10 kpc) from the nucleus of the S0 Seyfert galaxy NGC 5252 (redshift $z$ = 0.0229; \citealt{2015ApJ...814....8K}). It has a \textit{Chandra} 0.5-8 keV X-ray luminosity of $\sim1.5 \times 10^{40}$ erg s$^{-1}$, which does not qualify it as an extreme ULX, and no significant signs of X-ray variability. However, it has some peculiar properties compared to other ULXs: (i) it has an optical counterpart that is clearly detected in ultraviolet and optical images obtained with the Sloan Digital Sky Survey (SDSS) and the \textit{Hubble} Space Telescope ($m_\mathrm{r} \sim$22 mag; \citealt{2015ApJ...814....8K}); (ii) its optical spectrum shows strong emission lines with a fairly small ($\sim$ 13 km s$^{-1}$) systematic velocity offset from that of the nucleus of the galaxy, revealing that the ULX is likely associated with NGC 5252 (\citealt{2015ApJ...814....8K}); (iii) the ULX is able to ionize the surrounding gas and to influence its kinematics, as revealed by the signs of gas rotation centered on the ULX (\citealt{2017ApJ...844L..21K}); (iv) it has a strong radio counterpart, with an average flux density in Very Large Array (VLA) observations of up to 0.3 arcsec resolution of 3.2 mJy and 1.4 mJy at 1.4 GHz and 8.4 GHz, respectively, and of 1.9 mJy at 4.9 GHz (\citealt{1994AJ....107.1227W}; \citealt{1995MNRAS.276.1262K}; \citealt{1995ApJ...450..559B}; see table 2 in \citealt{2017MNRAS.464L..70Y}). The ULX appears unresolved in all these observations, as well as in 1.6 GHz Multi-Element Radio Linked Interferometer Network (MERLIN) observations (\citealt{2001MNRAS.327..369T}) and in 1.6 GHz European VLBI Network (EVN) observations with a resolution of 3 milli-arcsec (mas; \citealt{2017MNRAS.464L..70Y}), and shows no evidence for variability of $>$ 10\% over 3 years at 1.4 GHz and 8.4 GHz (\citealt{2015ApJ...814....8K}). All the above led to the conclusion that CXO J133815.6+043255 is not powered by a stellar-mass BH (\citealt{2015ApJ...814....8K,2017ApJ...844L..21K}).

In this Letter we report Very Long Baseline Array (VLBA) radio observations of the ULX CXO J133815.6+043255 at 4.4 GHz and 7.6 GHz that resolve, for the first time, its radio emission. The results make the blazar nature of the ULX very unlikely, in agreement with optical studies, and provide a more robust estimate of the ULX BH mass than in previous works, placing it in the realm of IMBHs. The observations, data reduction, and results obtained are described in Section~\ref{observations} and discussed in Section~\ref{discussion}. Final conclusions and open issues are provided in Sect.~\ref{conclusions}. Throughout the paper we adopt a $\Lambda$CDM cosmology with parameters $H_{0}=71$ km s$^{-1}$ Mpc$^{-1}$, $\Omega_{\Lambda}=0.73$ and $\Omega_{m}=0.27$, which yields a luminosity distance for NGC 5252 of 98.4 Mpc.

\section{Observations and data reduction}
\label{observations}
The ULX CXO J133815.6+043255 was observed with the VLBA for eight hours on 2016 March 23 (project BL0226) simultaneously at 4.4 GHz and 7.6 GHz in order to provide spectral index information. Ten antennas participated in the observations: Brewster, Fort Davis, Hancock, Kitt Peak, Los Alamos, Mauna Kea, North Liberty, Owens Valley, Pie Town, and Saint Croix. The observations were performed in phase-reference mode, alternating between $\sim$5 min on the target source and $\sim$1 min on a nearby phase calibrator (J1330+071). The bright radio source 3C279 was also observed as fringe finder and bandpass calibrator. Each of the two frequencies were recorded at a rate of 1024 Mbps in dual circular polarization, using a bandwidth of 128 MHz split into 258 spectral channels, and were correlated in Socorro (New Mexico) with an averaging time of 2 s. 

The correlated visibility data were split into two uv datasets, centered at 4.4 GHz and 7.6 GHz, and independently calibrated in amplitude (based on system temperatures and antenna sensitivities) and fringe-fitted using the NRAO package Astronomical Imaging Processing Software (\textsc{AIPS}) software. Sampling-based calibration adjustments determined with the task \textsc{ACCOR} and ionospheric correction were also applied at both frequencies. We consider a systematic uncertainty of 5\% in the flux calibration. The imaging was performed in \textsc{AIPS} using \textsc{CLEAN} deconvolution and the natural weighting of the data. The resulting 4.4 GHz image has a root mean square (r.m.s.) noise of 26.4 $\mu$Jy beam$^{-1}$ and a synthesized beam size of 4.94 $\times$ 1.95 mas$^{2}$. To recover any diffuse emission we also tried imaging without natural weighting and the robust parameter set to 0; however, the extent of the 4.4 GHz emission remains the same as when using natural weighting. To obtain the radio map at 7.6 GHz we used the same beam size as that of the 4.4 GHz radio map in order to derive the spectral index. The resulting rms at 7.6 GHz is of 23.5 $\mu$Jy beam$^{-1}$. The imaged data were fitted with two-dimensional elliptical Gaussians with the \textsc{AIPS} task \textsc{IMFIT}. For the phase calibrator J1330+071, we measured an integrated flux density of $\sim$0.1 Jy at 4.4 GHz and 7.6 GHz. For the target CXO J133815.6+043255, the distance between the two peaks at 4.4 GHz was measured from the lowest 4$\sigma$ contours using the \textsc{AIPS} task \textsc{TVDIST}. The final images were produced using the \textsc{CASA}\footnote{\textsc{COMMON\ ASTRONOMY\ SOFTWARE\ APPLICATIONS}} software.  

\begin{figure}
\includegraphics[width=0.43\textwidth]{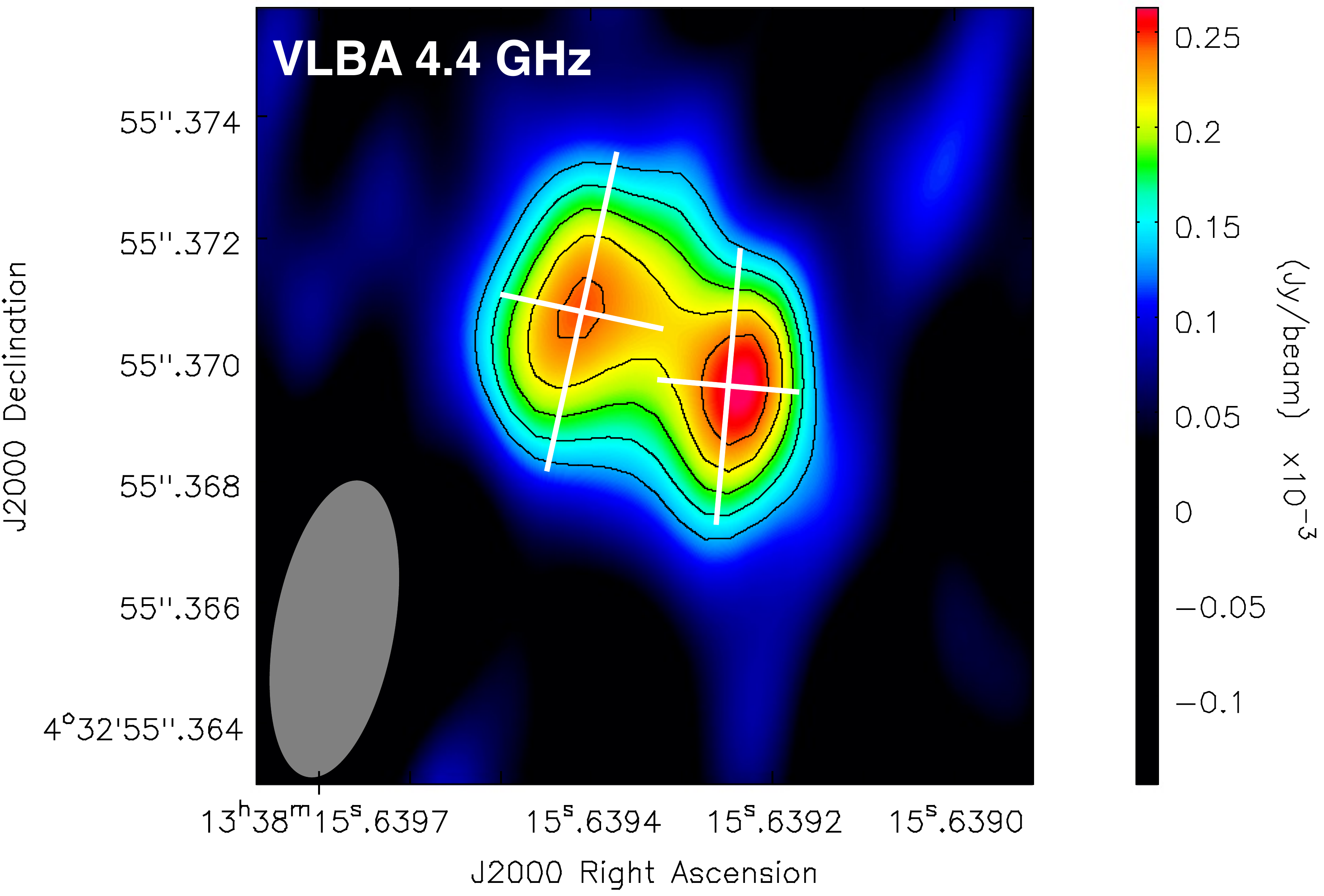}
\includegraphics[width=0.43\textwidth]{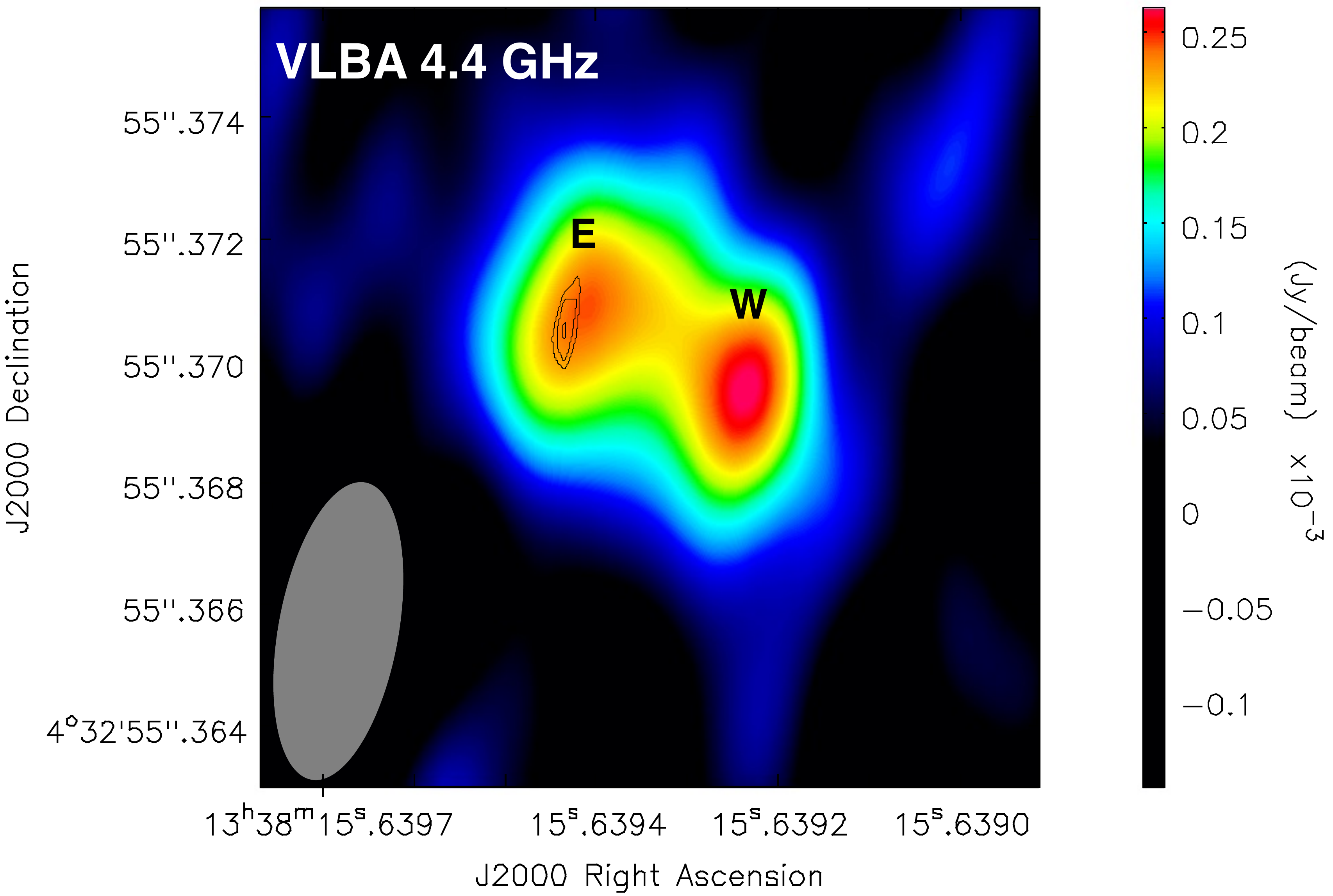}
\protect\caption[figure]{VLBA image of the ULX CXO J133815.6+043255 at 4.4 GHz. The synthesized beam size is 4.94 $\times$ 1.95 mas$^{2}$ with the major axis oriented at a position angle P.A.=$169.7^{\circ}$. \textbf{Top}: The 4.4 GHz contours are plotted as (-3,4,5,6,7,8,9) $\times$ the off-source r.m.s. noise of 26.4 $\mu$Jy beam$^{-1}$. The position and size of the components derived from the two-dimensional Gaussian fitting are marked with white crosses. The peak flux density of the east component is 0.24 mJy beam$^{-1}$, that of the west component is 0.26 mJy beam$^{-1}$. \textbf{Bottom}. The 7.6 GHz contours are plotted as (-3, 4, 4.5, 5) $\times$ the off-source r.m.s. noise of 23.5 $\mu$Jy beam$^{-1}$. The peak flux density of the 7.6 GHz emission is 0.12 mJy beam$^{-1}$ and is coincident with the 4.4 GHz east peak within the positional errors. North is up and east is left. The east and west 4.4 GHz peaks are marked as 'E' and 'W', respectively.}
\label{figure}
\end{figure}

\subsection{VLBA detection of the ULX CXO J133815.6+043255}
\label{detection}
The ULX CXO J133815.6+043255 is detected with the VLBA at 4.4 GHz and 7.6 GHz. At 4.4 GHz two peaks of radio emission separated by 2.9 mas (1.4 pc at a distance of 98.4 Mpc) are detected, each at a signal-to-noise ratio S/N$\sim$9-10 (see Fig.~\ref{figure}). The extended radio structure is oriented NE-SW. The fit of a double (one per peak) two-dimensional elliptical Gaussian gives a flux density for the east component of 0.36 $\pm$ 0.07 mJy and for the west one of 0.29 $\pm$ 0.06 mJy (see Table~\ref{gaussianfitting}), where the errors have been derived as the quadratic sum of the uncertainty resulting from the Gaussian fitting and the 5\% systematic uncertainty on the flux densities. The whole structure has a total integrated flux density of 0.66 $\pm$ 0.09 mJy. The eastern component has a deconvolved size of 2.2 $\times$ 1.7 mas$^{2}$ (1.0 $\times$ 0.8 pc$^{2}$) oriented at a P.A. of 139$^{\circ}$. The western one is oriented at a P.A. of 155$^{\circ}$ and is resolved only on its major axis, hence its size of 1.5 mas (0.7 pc) should be taken as an upper limit. According to these results, the two components have a brightness temperature $T_\mathrm{B} > 5 \times 10^{6}$ K, indicating that the emission is non-thermal.

At 7.6 GHz a single component that is not resolved is detected at S/N$\sim$5 (Fig.~\ref{figure}, bottom). Its peak flux density is 0.12 $\pm$ 0.03 mJy beam$^{-1}$. The 7.6 GHz detection is spatially coincident within the positional errors\footnote{The total positional error of each of the detected components is estimated, at each frequency, as the quadratic sum of the positional error of that component in the phase-referenced map, the positional error of the phase-reference calibrator, and the error of phase referencing due to ionospheric effects.} with the eastern component at 4.4 GHz (Fig.~\ref{figure}, bottom). This allows us to derive the spectral index for the eastern component, finding a steep value $\alpha = -2.0 \pm 0.1$ (where $S_{\nu} \propto \nu^{\alpha}$). For the west component we derive a 5$\sigma$ upper limit on its 7.6 GHz peak flux density of 0.2 $\mu$Jy beam$^{-1}$ from the r.m.s at the 4.4 GHz position. This yields an upper limit on its spectral index of $\alpha = -0.6$.

\begin{table*}
\begin{minipage}{\textwidth}
\centering
\caption{Results of the elliptical Gaussian fitting of the ULX CXO J133815.6+043255. Column designation:~(1) Component detected at S/N $>$ 5, (2-4) coordinates and positional error, (5) integrated flux, (6) peak flux, (7-8) angular and (projected) physical size, (9) brightness temperature.}
\label{gaussianfitting}
\begin{tabular}{lccccccc}
\hline
\hline 
  Freq.  		& 	RA						&	  DEC							& 	Pos. Error	 &  total 			& 	peak 			& Size								& $T_\mathrm{B}$ 			\\
 (GHz)  		& 	(J2000)	 				& 	  (J2000)							&	(mas)	 & (mJy)			& (mJy beam$^{-1}$)	& (mas)								&  (K)  					\\
\hline
4.4 east    	& 	13$^{h}$38$^{m}$15$^{s}$.639	 &  +04$^{\circ}$32\arcmin 55\arcsec.3708	& 	0.9		& 0.36 $\pm$ 0.07  & 0.24 $\pm$ 0.03		 & 2.2 $\times$ 1.7 (1.0 $\times$ 0.8 pc$^{2}$)	&  $6.3 \times 10^{6}$		\\
4.4 west 		   &	13$^{h}$38$^{m}$15$^{s}$.639	 &  +04$^{\circ}$32\arcmin 55\arcsec.3696	& 	0.8		& 0.29 $\pm$ 0.06  & 0.26 $\pm$ 0.03		&  $<$1.5 ($<$0.7 pc)					&  $\geq 5.5 \times 10^{6}$	\\
7.6 east   		& 	13$^{h}$38$^{m}$15$^{s}$.639	 &  +04$^{\circ}$32\arcmin 55\arcsec.3706	& 	1.0		&      -		      &  0.12 $\pm$ 0.03		& $<$2.1	($<$1.0 pc)					& 		 -				\\
\hline
\hline
\end{tabular}
\end{minipage}
\end{table*}

\section{Discussion}
\label{discussion}
\subsection{The extended radio jet of the ULX}
The 4.4 GHz VLBA observations of ULX CXO J133815.6+043255 reveal the first detection of a ULX pc-scale jet resolved into two components. All previous VLBI observations of ULXs report, as far as we know, either compact or slightly extended radio cores (e.g., \citealt{2011AN....332..379M}; \citealt{2013MNRAS.436.3128M,2014ApJ...785..121M,2015MNRAS.448.1893M}; \citealt{2015MNRAS.446.3268C,2015MNRAS.452...24C}), including the 1.6 GHz EVN observations of the ULX CXO J133815.6+043255 (\citealt{2017MNRAS.464L..70Y}). These EVN observations showed the detection of a compact radio component with a flux density of 1.8 $\pm$ 0.1 mJy and a flat spectral index $\alpha \sim$ -0.1, which \cite{2017MNRAS.464L..70Y} identified as the radio core of CXO J133815.6+043255. The EVN compact radio structure is however resolved by our higher-resolution 4.4 GHz VLBA observations into two components with a total flux density of 0.66 $\pm$ 0.09 mJy, indicating that the 1.6 GHz radio detection has some diffuse jet emission and that it cannot be ascribed to the core. This is reinforced by the steep $\alpha \leq$-1.6 and $\alpha \leq$-1.9 (considered as upper limits given the different beam resolution and non-simultaneity of the observations) found for the east and west components, respectively, when using the 1.6 GHz flux density of \cite{2017MNRAS.464L..70Y} and the 4.4 GHz VLBA fluxes to derive the spectral indexes. 

From the 4.4 and 7.6 GHz VLBA detections we find that the east component has a steep spectral index ($\alpha = -2.0 \pm 0.1$) and that it is resolved with a size of 1.0 $\times$ 0.8 pc$^{2}$ at 4.4 GHz, which precludes its identification as the radio core of CXO J133815.6+043255. We note that its $T_\mathrm{B} = 6.3 \times 10^{6}$ K is low compared to the equipartition value of compact relativistic jets (T$_\mathrm{B, eq.} \simeq 5 \times 10^{10}$ K; \citealt{1994ApJ...426...51R}). Given that the Doppler factor is $\sim T_\mathrm{B}/T_\mathrm{B, eq.}$, no significant Doppler-boosting is present in the east radio component. For the west one, only a lower limit on $T_\mathrm{B}$ can be derived given its partial (only major axis) resolved structure; hence the presence of Doppler-boosting cannot be ruled out here. Given that this component is not firmly resolved and has a flatter spectral index ($\alpha \leq -0.6$) than the east one, the radio core is more likely to be located here. For the purposes of investigating the nature of CXO J133815.6+043255, in the next section we consider the flux of the west detection as an upper limit to that of the radio core. 

\subsection{The nature of the ULX }
The resolved jet structure and steep spectra revealed by the VLBA observations suggest that the ULX is not a background blazar, as the radio emission of most blazars is dominated by an unresolved flat-spectrum core. This is in agreement with optical spectroscopic studies, which locate the ULX in NGC 5252, and with the lack of significant radio variability over 3 years (\citealt{2015ApJ...814....8K,2017ApJ...844L..21K}). It should be noted though that, when observed at low frequencies, some blazars can show extended radio emission (e.g. \citealt{2010ApJ...710..764K}). From the radio emission standpoint the detected VLBA emission could be also consistent with that of compact steep spectrum (CSS) sources, which are young radio sources with steep-spectrum small-scale jets (\citealt{1998PASP..110..493O}). However, CSS sources have typically strong optical emission lines (e.g. \citealt{1997A&A...326..130M}; \citealt{2016A&ARv..24...10T}), hence the nature of CXO J133815.6+043255 as a CSS is ruled out by its optical spectrum showing that it belongs to NGC 5252 (\citealt{2015ApJ...814....8K}). 

NGC 5252 seems to have undergone a past interaction, as evidenced by the finding of a small-scale half-spiral of dust near the nucleus of the galaxy and of a kinematical decoupling between the stars and the gas (\citealt{1998ApJ...505..159M}; \citealt{2015AJ....149..155K}). This, together with the size of the optical counterpart of CXO J133815.6+043255 ($\lesssim$ 46 pc; \citealt{2015ApJ...814....8K}) being consistent with that of ultracompact dwarf galaxies, suggests that CXO J133815.6+043255 is the nucleus of a dwarf galaxy that was accreted by NGC 5252 (\citealt{2015ApJ...814....8K,2017ApJ...844L..21K}). In this scenario, the ULX could be either an AGN powered by an IMBH or a low-luminosity AGN (LLAGN). 

LLAGN host supermassive BHs with masses $> 10^{6}$ M$_{\odot}$, have typically X-ray luminosities $\sim10^{40-41}$ erg s$^{-1}$ (\citealt{2008ARA&A..46..475H}) and their nuclear radio emission is usually associated with unresolved cores (e.g., \citealt{2001ApJ...558..561U}; \citealt{2005A&A...435..521N}; \citealt{2018MNRAS.476.3478B}; \citealt{2018arXiv180506696S}). CXO J133815.6+043255 shows [OIII] and X-ray luminosities of 10$^{39.7}$ and 10$^{40.2}$ erg s$^{-1}$, respectively (\citealt{2015ApJ...814....8K}), consistent with those of LLAGN. Because of this and based on the finding of compact radio emission with a flat spectrum, \cite{2017MNRAS.464L..70Y} argued that CXO J133815.6+043255 is a LLAGN. However, when observed with sufficient angular resolution and sensitivity, LLAGN can show resolved pc-scale radio emission (e.g., \citealt{2014ApJ...787...62M}; \citealt{2018MNRAS.476.3478B}). This could be the case of CXO J133815.6+043255, for which we find that the pc-scale radio jet is resolved into two components. Given their size and radio spectral index, we ascribe the east component to a radio lobe and the west one to the radio core. 

To probe the nature of CXO J133815.6+043255, we compute the $R_\mathrm{X}$ ratio of 5 GHz radio luminosity to 2-10 keV X-ray luminosity (\citealt{2003ApJ...583..145T}). LLAGN have typically $-3.8 <$ log $R_\mathrm{X} < -2.8$, X-ray binaries log $R_\mathrm{X} < -5.3$, supernova remnants log $R_\mathrm{X} \sim -2$, and IMBHs $-5.3 <$ log $R_\mathrm{X} < -3.8$ (\citealt{2013MNRAS.436.1546M,2013MNRAS.436.2454M}). Using the 4.4 GHz flux density of the western component we find log $R_\mathrm{X}$ = -3, consistent with LLAGN. Note though that since the western component is slightly resolved, its flux density should be taken as an upper limit to the core radio emission and so the value of $R_\mathrm{X}$. The same result would be obtained when considering the flux of the east component as an upper limit to the core emission. 

Those BHs accreting at sub-Eddington rates and in a low/hard X-ray state are found to follow an empirical correlation, supported by theoretical models of accretion, that relates their nuclear X-ray luminosity with their core radio luminosity and BH mass (e.g., \citealt{2004A&A...414..895F}; \citealt{2006A&A...456..439K}; \citealt{2009ApJ...706..404G}; \citealt{2012MNRAS.419..267P}; \citealt{2018arXiv180506696S}; see \citealt{2018MNRAS.474.1342M} for a brief review). Using this fundamental plane of BH accretion, \cite{2017MNRAS.464L..70Y} estimated a BH mass for CXO J133815.6+043255 of $\sim10^{9}$ M$_{\odot}$, which is unreasonably large for an AGN in a dwarf galaxy. The finding that their EVN radio emission is resolved when using higher-resolution VLBA observations indicates that the BH mass of $\sim10^{9}$ M$_{\odot}$ should be taken as a very rough upper limit to the ULX BH mass. From the 4.4 GHz radio luminosity of the western VLBA component ($L_\mathrm{R} = 1.5 \times 10^{37}$ erg s$^{-1}$), which we consider as an upper limit to the core radio luminosity, and the ULX 2-10 keV X-ray luminosity ($L_\mathrm{X} =  1.2 \times 10^{40}$ erg s$^{-1}$; \citealt{2015ApJ...814....8K}), we estimate an upper limit on the ULX BH mass of $M_\mathrm{BH}\lesssim 2  \times 10^{6}$ M$_{\odot}$ when using the most recent and refined version of the fundamental plane of BH accretion (\citealt{2018arXiv180506696S}):\\
\begin{equation}
\begin{split}
\mathrm{log} L_\mathrm{R} = (0.48 \pm 0.04) \mathrm{log} L_\mathrm{X} + (0.79 \pm 0.03) \mathrm{log} M_\mathrm{BH} + 11.71 
\end{split}
\end{equation}
The fundamental plane of \cite{2018arXiv180506696S} has been derived from an homogeneous sample of LLAGN with core radio emission based on uniform resolution and sensitivity VLA observations at 15 GHz, while previous correlations were derived from radio flux densities that could possibly trace nuclear jet luminosities and not only core luminosities (\citealt{2018arXiv180506696S}). Using the \cite{2009ApJ...706..404G} correlation we obtain $M_\mathrm{BH}\lesssim 10^{7}$ M$_{\odot}$, while using those of \cite{2006A&A...456..439K} and \cite{2012MNRAS.419..267P} we estimate $M_\mathrm{BH} \lesssim 6  \times 10^{8}$ M$_{\odot}$. Similar upper limits would be obtained when considering the flux of the east component as an upper limit to the core emission.

The upper limit on the ULX BH mass of $\lesssim2 \times 10^{6}$ M$_{\odot}$ is in agreement with the dynamical mass $M_\mathrm{dyn} = 4 \times 10^{7}$ M$_{\odot}$ derived by \cite{2017ApJ...844L..21K} and with the lower limit of $10^{3.5}$ M$_{\odot}$ obtained assuming that the bolometric luminosity does not exceed the Eddington luminosity (\citealt{2015ApJ...814....8K}). It is also consistent with the $M_\mathrm{BH}\lesssim 10^{6}$ M$_{\odot}$ of most of the low-mass AGN found in dwarf galaxies (e.g., \citealt{2003ApJ...588L..13F}; \citealt{2004ApJ...607...90B}; \citealt{2004ApJ...610..722G,2007ApJ...670...92G}; \citealt{2013ApJ...775..116R,2014ApJ...787L..30R}; \citealt{2015ApJ...809L..14B}; \citealt{2015ApJ...798...38S}; \citealt{2017ApJ...836..237N}; \citealt{2016ApJ...817...20M,2018MNRAS.478.2576M}), suggesting that the ULX hosts also a low-mass AGN. 

Low-mass AGN in dwarf galaxies are typically found to accrete at near-Eddington rates. Detections of sub-Eddington accreting AGN in dwarf galaxies showing radio jets are scarce (e.g., NGC 4395, \citealt{2006ApJ...646L..95W}; NGC 404, \citealt{2012ApJ...753..103N}; only 1 out of 19 objects in \citealt{2006ApJ...636...56G}; 3 out of 40 sources in \citealt{2018MNRAS.478.2576M}); hence the finding of an additional source such as CXO J133815.6+043255 is very significant. Assuming a BH mass of $2 \times 10^{6}$ M$_{\odot}$ and a bolometric luminosity $L_\mathrm{bol} = 8 \times 10^{41}$ erg s$^{-1}$ (derived from the [OIII] luminosity adopting the conversion $L_\mathrm{bol}/L_\mathrm{[OIII]}\sim162$; \citealt{2012MNRAS.426.2703S}; \citealt{2015ApJ...814....8K}), we derive an Eddington ratio for CXO J133815.6+043255 of $3 \times 10^{-3}$. Using the higher BH mass limit of $10^{7}$ M$_{\odot}$ or $6  \times 10^{8}$ M$_{\odot}$ derived from other fundamental correlations the ULX would have an Eddington ratio $\sim 10^{-7}-10^{-5}$. This would produce a ionization parameter for the optical emission lines typical of low-ionization nuclear emission-line regions (LINERs), which have [OIII]/H${\beta} < 3$ (\citealt{1997ApJS..112..315H}). Instead, CXO J133815.6+043255 is observed to have [OIII]/H${\beta} \sim 9$ typical of Seyfert galaxies (\citealt{2015ApJ...814....8K}). We thus conclude that the ULX is most likely to host an IMBH with $10^{3.5} < M_\mathrm{BH}$ (M$_{\odot}$) $\lesssim 2 \times 10^{6}$ rather than a supermassive BH. 

\section{Conclusions and open issues}
\label{conclusions}
IMBHs are thought to be the local relics of the early Universe supermassive BH progenitors. Finding observational evidence of their existence is thus of paramount importance for understanding how supermassive BHs grow. Most IMBH candidates are found as low-mass AGN with no radio emission and near- to super-Eddington accretion rates, as expected from simulations in which early BH growth proceeds through short high-accretion-rate phases (e.g., \citealt{2005ApJ...633..624V}; \citealt{2016MNRAS.458.3047P}; \citealt{2017MNRAS.472L.109A}). The finding of low-mass AGN with extended radio jets and sub-Eddington accretion rates such as CXO J133815.6+043255 has thus important implications for cosmological models. 

Unlike most low-mass AGN, CXO J133815.6+043255 is a ULX located 10 kpc away from the center of its host galaxy. This suggests that it is the nucleus of a dwarf galaxy that was stripped in the course of a minor merger. Numerical simulations show that BH growth in dwarf galaxies can be triggered by minor mergers; however, evidence of AGN-triggered activity in a dwarf galaxy is scarce (\citealt{2013MNRAS.435.2335B}; \citealt{2017ApJ...836..183S}). CXO J133815.6+043255 could be a new case of a BH that became active in the course of a minor merger event. 

The ULX CXO J133815.6+043255 is also one of the few low-mass AGN known able to ionize the gas that surrounds it (\citealt{2017ApJ...844L..21K}). This incidence of AGN feedback in dwarf galaxies is an issue of major debate. Most numerical simulations maintain that supernova feedback hampers BH growth and thus the impact of AGN feedback in low-mass galaxies (e.g., \citealt{2015MNRAS.452.1502D}; \citealt{2017MNRAS.465...32B}; \citealt{2017MNRAS.472L.109A}; \citealt{2017MNRAS.468.3935H}), others that AGN feedback is the one that has the biggest impact on the stellar populations of dwarf galaxies (\citealt{2016MNRAS.463.2986S}; \citealt{2018MNRAS.473.5698D}). Observationally the results are also controversial: \cite{2018ApJ...855L..20M} find support for the supernova feedback scenario, while \cite{2018MNRAS.476..979P} report that AGN feedback could regulate star formation in dwarf galaxies. CXO J133815.6+043255 seems to bolster this possibility. Further observational studies are needed to clarify the roles that supernova and AGN feedback play in regulating BH growth in dwarf galaxies.

\section*{Acknowledgments}
The authors thank the anonymous referee for insightful comments. M.M. acknowledges support from the Spanish Juan de la Cierva program (IJCI-2015-23944). M.K. was supported by the Basic Science Research Program through the National Research Foundation of Korea (NRF) funded by the Ministry of Science, ICT \& Future Planning (No. NRF-2017R1C1B2002879). L.C.H. was supported by the National Key R\&D Program of China (2016YFA0400702) and the National Science Foundation of China (11473002, 11721303).


\bibliographystyle{mnras}
\bibliography{/Users/mmezcua/Documents/referencesALL}

\begin{thebibliography}{}
\makeatletter
\relax
\def\mn@urlcharsother{\let\do\@makeother \do\$\do\&\do\#\do\^\do\_\do\%\do\~}
\def\mn@doi{\begingroup\mn@urlcharsother \@ifnextchar [ {\mn@doi@}
  {\mn@doi@[]}}
\def\mn@doi@[#1]#2{\def\@tempa{#1}\ifx\@tempa\@empty \href
  {http://dx.doi.org/#2} {doi:#2}\else \href {http://dx.doi.org/#2} {#1}\fi
  \endgroup}
\def\mn@eprint#1#2{\mn@eprint@#1:#2::\@nil}
\def\mn@eprint@arXiv#1{\href {http://arxiv.org/abs/#1} {{\tt arXiv:#1}}}
\def\mn@eprint@dblp#1{\href {http://dblp.uni-trier.de/rec/bibtex/#1.xml}
  {dblp:#1}}
\def\mn@eprint@#1:#2:#3:#4\@nil{\def\@tempa {#1}\def\@tempb {#2}\def\@tempc
  {#3}\ifx \@tempc \@empty \let \@tempc \@tempb \let \@tempb \@tempa \fi \ifx
  \@tempb \@empty \def\@tempb {arXiv}\fi \@ifundefined
  {mn@eprint@\@tempb}{\@tempb:\@tempc}{\expandafter \expandafter \csname
  mn@eprint@\@tempb\endcsname \expandafter{\@tempc}}}

\bibitem[\protect\citeauthoryear{{Angl{\'e}s-Alc{\'a}zar},
  {Faucher-Gigu{\`e}re}, {Quataert}, {Hopkins}, {Feldmann}, {Torrey}, {Wetzel}
  \& {Kere{\v s}}}{{Angl{\'e}s-Alc{\'a}zar} et~al.}{2017}]{2017MNRAS.472L.109A}
{Angl{\'e}s-Alc{\'a}zar} D.,  {Faucher-Gigu{\`e}re} C.-A.,  {Quataert} E.,
  {Hopkins} P.~F.,  {Feldmann} R.,  {Torrey} P.,  {Wetzel} A.,   {Kere{\v s}}
  D.,  2017, \mn@doi [\mnras] {10.1093/mnrasl/slx161}, \href
  {http://adsabs.harvard.edu/abs/2017MNRAS.472L.109A} {472, L109}

\bibitem[\protect\citeauthoryear{{Ba{\~n}ados} et~al.,}{{Ba{\~n}ados}
  et~al.}{2018}]{2018Natur.553..473B}
{Ba{\~n}ados} E.,  et~al., 2018, \mn@doi [\nat] {10.1038/nature25180}, \href
  {http://adsabs.harvard.edu/abs/2018Natur.553..473B} {553, 473}

\bibitem[\protect\citeauthoryear{{Bachetti} et~al.,}{{Bachetti}
  et~al.}{2014}]{2014Natur.514..202B}
{Bachetti} M.,  et~al., 2014, \mn@doi [\nat] {10.1038/nature13791}, \href
  {http://adsabs.harvard.edu/abs/2014Natur.514..202B} {514, 202}

\bibitem[\protect\citeauthoryear{{Baldassare}, {Reines}, {Gallo}  \&
  {Greene}}{{Baldassare} et~al.}{2015}]{2015ApJ...809L..14B}
{Baldassare} V.~F.,  {Reines} A.~E.,  {Gallo} E.,   {Greene} J.~E.,  2015,
  \mn@doi [\apjl] {10.1088/2041-8205/809/1/L14}, \href
  {http://adsabs.harvard.edu/abs/2015ApJ...809L..14B} {809, L14}

\bibitem[\protect\citeauthoryear{{Baldassare}, {Reines}, {Gallo}  \&
  {Greene}}{{Baldassare} et~al.}{2017}]{2017ApJ...836...20B}
{Baldassare} V.~F.,  {Reines} A.~E.,  {Gallo} E.,   {Greene} J.~E.,  2017,
  \mn@doi [\apj] {10.3847/1538-4357/836/1/20}, \href
  {http://adsabs.harvard.edu/abs/2017ApJ...836...20B} {836, 20}

\bibitem[\protect\citeauthoryear{{Baldi} et~al.,}{{Baldi}
  et~al.}{2018}]{2018MNRAS.476.3478B}
{Baldi} R.~D.,  et~al., 2018, \mn@doi [\mnras] {10.1093/mnras/sty342}, \href
  {http://adsabs.harvard.edu/abs/2018MNRAS.476.3478B} {476, 3478}

\bibitem[\protect\citeauthoryear{{Barth}, {Ho}, {Rutledge}  \&
  {Sargent}}{{Barth} et~al.}{2004}]{2004ApJ...607...90B}
{Barth} A.~J.,  {Ho} L.~C.,  {Rutledge} R.~E.,   {Sargent} W.~L.~W.,  2004,
  \mn@doi [\apj] {10.1086/383302}, \href
  {http://adsabs.harvard.edu/abs/2004ApJ...607...90B} {607, 90}

\bibitem[\protect\citeauthoryear{{Becker}, {White}  \& {Helfand}}{{Becker}
  et~al.}{1995}]{1995ApJ...450..559B}
{Becker} R.~H.,  {White} R.~L.,   {Helfand} D.~J.,  1995, \mn@doi [\apj]
  {10.1086/176166}, \href {http://adsabs.harvard.edu/abs/1995ApJ...450..559B}
  {450, 559}

\bibitem[\protect\citeauthoryear{{Bianchi}, {Piconcelli}, {P{\'e}rez-Torres},
  {Fiore}, {La Franca}, {Mathur}  \& {Matt}}{{Bianchi}
  et~al.}{2013}]{2013MNRAS.435.2335B}
{Bianchi} S.,  {Piconcelli} E.,  {P{\'e}rez-Torres} M.~{\'A}.,  {Fiore} F.,
  {La Franca} F.,  {Mathur} S.,   {Matt} G.,  2013, \mn@doi [\mnras]
  {10.1093/mnras/stt1459}, \href
  {http://adsabs.harvard.edu/abs/2013MNRAS.435.2335B} {435, 2335}

\bibitem[\protect\citeauthoryear{{Bower}, {Schaye}, {Frenk}, {Theuns},
  {Schaller}, {Crain}  \& {McAlpine}}{{Bower}
  et~al.}{2017}]{2017MNRAS.465...32B}
{Bower} R.~G.,  {Schaye} J.,  {Frenk} C.~S.,  {Theuns} T.,  {Schaller} M.,
  {Crain} R.~A.,   {McAlpine} S.,  2017, \mn@doi [\mnras]
  {10.1093/mnras/stw2735}, \href
  {http://adsabs.harvard.edu/abs/2017MNRAS.465...32B} {465, 32}

\bibitem[\protect\citeauthoryear{{Cseh} et~al.,}{{Cseh}
  et~al.}{2015a}]{2015MNRAS.446.3268C}
{Cseh} D.,  et~al., 2015a, \mn@doi [\mnras] {10.1093/mnras/stu2363}, \href
  {http://adsabs.harvard.edu/abs/2015MNRAS.446.3268C} {446, 3268}

\bibitem[\protect\citeauthoryear{{Cseh} et~al.,}{{Cseh}
  et~al.}{2015b}]{2015MNRAS.452...24C}
{Cseh} D.,  et~al., 2015b, \mn@doi [\mnras] {10.1093/mnras/stv1308}, \href
  {http://adsabs.harvard.edu/abs/2015MNRAS.452...24C} {452, 24}

\bibitem[\protect\citeauthoryear{{Dashyan}, {Silk}, {Mamon}, {Dubois}  \&
  {Hartwig}}{{Dashyan} et~al.}{2018}]{2018MNRAS.473.5698D}
{Dashyan} G.,  {Silk} J.,  {Mamon} G.~A.,  {Dubois} Y.,   {Hartwig} T.,  2018,
  \mn@doi [\mnras] {10.1093/mnras/stx2716}, \href
  {http://adsabs.harvard.edu/abs/2018MNRAS.473.5698D} {473, 5698}

\bibitem[\protect\citeauthoryear{{Davis}, {Narayan}, {Zhu}, {Barret},
  {Farrell}, {Godet}, {Servillat}  \& {Webb}}{{Davis}
  et~al.}{2011}]{2011ApJ...734..111D}
{Davis} S.~W.,  {Narayan} R.,  {Zhu} Y.,  {Barret} D.,  {Farrell} S.~A.,
  {Godet} O.,  {Servillat} M.,   {Webb} N.~A.,  2011, \mn@doi [\apj]
  {10.1088/0004-637X/734/2/111}, \href
  {http://adsabs.harvard.edu/abs/2011ApJ...734..111D} {734, 111}

\bibitem[\protect\citeauthoryear{{Dubois}, {Volonteri}, {Silk}, {Devriendt},
  {Slyz}  \& {Teyssier}}{{Dubois} et~al.}{2015}]{2015MNRAS.452.1502D}
{Dubois} Y.,  {Volonteri} M.,  {Silk} J.,  {Devriendt} J.,  {Slyz} A.,
  {Teyssier} R.,  2015, \mn@doi [\mnras] {10.1093/mnras/stv1416}, \href
  {http://adsabs.harvard.edu/abs/2015MNRAS.452.1502D} {452, 1502}

\bibitem[\protect\citeauthoryear{{Falcke}, {K{\"o}rding}  \&
  {Markoff}}{{Falcke} et~al.}{2004}]{2004A&A...414..895F}
{Falcke} H.,  {K{\"o}rding} E.,   {Markoff} S.,  2004, \mn@doi [\aap]
  {10.1051/0004-6361:20031683}, \href
  {http://adsabs.harvard.edu/abs/2004A%26A...414..895F} {414, 895}

\bibitem[\protect\citeauthoryear{{Farrell}, {Webb}, {Barret}, {Godet}  \&
  {Rodrigues}}{{Farrell} et~al.}{2009}]{2009Natur.460...73F}
{Farrell} S.~A.,  {Webb} N.~A.,  {Barret} D.,  {Godet} O.,   {Rodrigues} J.~M.,
   2009, \mn@doi [\nat] {10.1038/nature08083}, \href
  {http://adsabs.harvard.edu/abs/2009Natur.460...73F} {460, 73}

\bibitem[\protect\citeauthoryear{{Filippenko} \& {Ho}}{{Filippenko} \&
  {Ho}}{2003}]{2003ApJ...588L..13F}
{Filippenko} A.~V.,  {Ho} L.~C.,  2003, \mn@doi [\apjl] {10.1086/375361}, \href
  {http://adsabs.harvard.edu/abs/2003ApJ...588L..13F} {588, L13}

\bibitem[\protect\citeauthoryear{{Greene} \& {Ho}}{{Greene} \&
  {Ho}}{2004}]{2004ApJ...610..722G}
{Greene} J.~E.,  {Ho} L.~C.,  2004, \mn@doi [\apj] {10.1086/421719}, \href
  {http://adsabs.harvard.edu/abs/2004ApJ...610..722G} {610, 722}

\bibitem[\protect\citeauthoryear{{Greene} \& {Ho}}{{Greene} \&
  {Ho}}{2007}]{2007ApJ...670...92G}
{Greene} J.~E.,  {Ho} L.~C.,  2007, \mn@doi [\apj] {10.1086/522082}, \href
  {http://adsabs.harvard.edu/abs/2007ApJ...670...92G} {670, 92}

\bibitem[\protect\citeauthoryear{{Greene}, {Ho}  \& {Ulvestad}}{{Greene}
  et~al.}{2006}]{2006ApJ...636...56G}
{Greene} J.~E.,  {Ho} L.~C.,   {Ulvestad} J.~S.,  2006, \mn@doi [\apj]
  {10.1086/497905}, \href {http://adsabs.harvard.edu/abs/2006ApJ...636...56G}
  {636, 56}

\bibitem[\protect\citeauthoryear{{G{\"u}ltekin}, {Cackett}, {Miller}, {Di
  Matteo}, {Markoff}  \& {Richstone}}{{G{\"u}ltekin}
  et~al.}{2009}]{2009ApJ...706..404G}
{G{\"u}ltekin} K.,  {Cackett} E.~M.,  {Miller} J.~M.,  {Di Matteo} T.,
  {Markoff} S.,   {Richstone} D.~O.,  2009, \mn@doi [\apj]
  {10.1088/0004-637X/706/1/404}, \href
  {http://adsabs.harvard.edu/abs/2009ApJ...706..404G} {706, 404}

\bibitem[\protect\citeauthoryear{{Habouzit}, {Volonteri}  \&
  {Dubois}}{{Habouzit} et~al.}{2017}]{2017MNRAS.468.3935H}
{Habouzit} M.,  {Volonteri} M.,   {Dubois} Y.,  2017, \mn@doi [\mnras]
  {10.1093/mnras/stx666}, \href
  {http://adsabs.harvard.edu/abs/2017MNRAS.468.3935H} {468, 3935}

\bibitem[\protect\citeauthoryear{{Ho}}{{Ho}}{2008}]{2008ARA&A..46..475H}
{Ho} L.~C.,  2008, \mn@doi [\araa] {10.1146/annurev.astro.45.051806.110546},
  \href {http://adsabs.harvard.edu/abs/2008ARA%26A..46..475H} {46, 475}

\bibitem[\protect\citeauthoryear{{Ho}, {Filippenko}  \& {Sargent}}{{Ho}
  et~al.}{1997}]{1997ApJS..112..315H}
{Ho} L.~C.,  {Filippenko} A.~V.,   {Sargent} W.~L.~W.,  1997, \mn@doi [\apjs]
  {10.1086/313041}, \href {http://adsabs.harvard.edu/abs/1997ApJS..112..315H}
  {112, 315}

\bibitem[\protect\citeauthoryear{{Israel} et~al.,}{{Israel}
  et~al.}{2017a}]{2017Sci...355..817I}
{Israel} G.~L.,  et~al., 2017a, \mn@doi [Science] {10.1126/science.aai8635},
  \href {http://adsabs.harvard.edu/abs/2017Sci...355..817I} {355, 817}

\bibitem[\protect\citeauthoryear{{Israel} et~al.,}{{Israel}
  et~al.}{2017b}]{2017MNRAS.466L..48I}
{Israel} G.~L.,  et~al., 2017b, \mn@doi [\mnras] {10.1093/mnrasl/slw218}, \href
  {http://adsabs.harvard.edu/abs/2017MNRAS.466L..48I} {466, L48}

\bibitem[\protect\citeauthoryear{{Kaaret}, {Prestwich}, {Zezas}, {Murray},
  {Kim}, {Kilgard}, {Schlegel}  \& {Ward}}{{Kaaret}
  et~al.}{2001}]{2001MNRAS.321L..29K}
{Kaaret} P.,  {Prestwich} A.~H.,  {Zezas} A.,  {Murray} S.~S.,  {Kim} D.-W.,
  {Kilgard} R.~E.,  {Schlegel} E.~M.,   {Ward} M.~J.,  2001, \mn@doi [\mnras]
  {10.1046/j.1365-8711.2001.04064.x}, \href
  {http://adsabs.harvard.edu/abs/2001MNRAS.321L..29K} {321, L29}

\bibitem[\protect\citeauthoryear{{Kaaret}, {Feng}  \& {Roberts}}{{Kaaret}
  et~al.}{2017}]{2017ARA&A..55..303K}
{Kaaret} P.,  {Feng} H.,   {Roberts} T.~P.,  2017, \mn@doi [\araa]
  {10.1146/annurev-astro-091916-055259}, \href
  {http://adsabs.harvard.edu/abs/2017ARA%26A..55..303K} {55, 303}

\bibitem[\protect\citeauthoryear{{Keel} et~al.,}{{Keel}
  et~al.}{2015}]{2015AJ....149..155K}
{Keel} W.~C.,  et~al., 2015, \mn@doi [\aj] {10.1088/0004-6256/149/5/155}, \href
  {http://adsabs.harvard.edu/abs/2015AJ....149..155K} {149, 155}

\bibitem[\protect\citeauthoryear{{Kharb}, {Lister}  \& {Cooper}}{{Kharb}
  et~al.}{2010}]{2010ApJ...710..764K}
{Kharb} P.,  {Lister} M.~L.,   {Cooper} N.~J.,  2010, \mn@doi [\apj]
  {10.1088/0004-637X/710/1/764}, \href
  {http://adsabs.harvard.edu/abs/2010ApJ...710..764K} {710, 764}

\bibitem[\protect\citeauthoryear{{Kim} et~al.,}{{Kim}
  et~al.}{2015}]{2015ApJ...814....8K}
{Kim} M.,  et~al., 2015, \mn@doi [\apj] {10.1088/0004-637X/814/1/8}, \href
  {http://adsabs.harvard.edu/abs/2015ApJ...814....8K} {814, 8}

\bibitem[\protect\citeauthoryear{{Kim}, {Ho}  \& {Im}}{{Kim}
  et~al.}{2017}]{2017ApJ...844L..21K}
{Kim} M.,  {Ho} L.~C.,   {Im} M.,  2017, \mn@doi [\apjl]
  {10.3847/2041-8213/aa7fe8}, \href
  {http://adsabs.harvard.edu/abs/2017ApJ...844L..21K} {844, L21}

\bibitem[\protect\citeauthoryear{{King} \& {Dehnen}}{{King} \&
  {Dehnen}}{2005}]{2005MNRAS.357..275K}
{King} A.~R.,  {Dehnen} W.,  2005, \mn@doi [\mnras]
  {10.1111/j.1365-2966.2005.08634.x}, \href
  {http://adsabs.harvard.edu/abs/2005MNRAS.357..275K} {357, 275}

\bibitem[\protect\citeauthoryear{{K{\"o}rding}, {Falcke}  \&
  {Corbel}}{{K{\"o}rding} et~al.}{2006}]{2006A&A...456..439K}
{K{\"o}rding} E.,  {Falcke} H.,   {Corbel} S.,  2006, \mn@doi [\aap]
  {10.1051/0004-6361:20054144}, \href
  {http://adsabs.harvard.edu/abs/2006A%26A...456..439K} {456, 439}

\bibitem[\protect\citeauthoryear{{Kukula}, {Pedlar}, {Baum}  \&
  {O'Dea}}{{Kukula} et~al.}{1995}]{1995MNRAS.276.1262K}
{Kukula} M.~J.,  {Pedlar} A.,  {Baum} S.~A.,   {O'Dea} C.~P.,  1995, \mn@doi
  [\mnras] {10.1093/mnras/276.4.1262}, \href
  {http://adsabs.harvard.edu/abs/1995MNRAS.276.1262K} {276, 1262}

\bibitem[\protect\citeauthoryear{{Liu}, {Bregman}, {Bai}, {Justham}  \&
  {Crowther}}{{Liu} et~al.}{2013}]{2013Natur.503..500L}
{Liu} J.-F.,  {Bregman} J.~N.,  {Bai} Y.,  {Justham} S.,   {Crowther} P.,
  2013, \mn@doi [\nat] {10.1038/nature12762}, \href
  {http://adsabs.harvard.edu/abs/2013Natur.503..500L} {503, 500}

\bibitem[\protect\citeauthoryear{{Mart{\'{\i}}n-Navarro} \&
  {Mezcua}}{{Mart{\'{\i}}n-Navarro} \& {Mezcua}}{2018}]{2018ApJ...855L..20M}
{Mart{\'{\i}}n-Navarro} I.,  {Mezcua} M.,  2018, \mn@doi [\apjl]
  {10.3847/2041-8213/aab103}, \href
  {http://adsabs.harvard.edu/abs/2018ApJ...855L..20M} {855, L20}

\bibitem[\protect\citeauthoryear{{McConnell}, {Ma}, {Gebhardt}, {Wright},
  {Murphy}, {Lauer}, {Graham}  \& {Richstone}}{{McConnell}
  et~al.}{2011}]{2011Natur.480..215M}
{McConnell} N.~J.,  {Ma} C.-P.,  {Gebhardt} K.,  {Wright} S.~A.,  {Murphy}
  J.~D.,  {Lauer} T.~R.,  {Graham} J.~R.,   {Richstone} D.~O.,  2011, \mn@doi
  [\nat] {10.1038/nature10636}, \href
  {http://adsabs.harvard.edu/abs/2011Natur.480..215M} {480, 215}

\bibitem[\protect\citeauthoryear{{Mezcua}}{{Mezcua}}{2017}]{2017IJMPD..2630021M}
{Mezcua} M.,  2017, \mn@doi [International Journal of Modern Physics D]
  {10.1142/S021827181730021X}, \href
  {http://adsabs.harvard.edu/abs/2017IJMPD..2630021M} {26, 1730021}

\bibitem[\protect\citeauthoryear{{Mezcua} \& {Lobanov}}{{Mezcua} \&
  {Lobanov}}{2011}]{2011AN....332..379M}
{Mezcua} M.,  {Lobanov} A.~P.,  2011, \mn@doi [Astron. Nachr.]
  {10.1002/asna.201011504}, \href
  {http://adsabs.harvard.edu/abs/2011AN....332..379M} {332, 379}

\bibitem[\protect\citeauthoryear{{Mezcua} \& {Prieto}}{{Mezcua} \&
  {Prieto}}{2014}]{2014ApJ...787...62M}
{Mezcua} M.,  {Prieto} M.~A.,  2014, \mn@doi [\apj]
  {10.1088/0004-637X/787/1/62}, \href
  {http://adsabs.harvard.edu/abs/2014ApJ...787...62M} {787, 62}

\bibitem[\protect\citeauthoryear{{Mezcua}, {Farrell}, {Gladstone}  \&
  {Lobanov}}{{Mezcua} et~al.}{2013a}]{2013MNRAS.436.1546M}
{Mezcua} M.,  {Farrell} S.~A.,  {Gladstone} J.~C.,   {Lobanov} A.~P.,  2013a,
  \mn@doi [\mnras] {10.1093/mnras/stt1674}, \href
  {http://adsabs.harvard.edu/abs/2013MNRAS.436.1546M} {436, 1546}

\bibitem[\protect\citeauthoryear{{Mezcua}, {Lobanov}  \&
  {Mart{\'{\i}}-Vidal}}{{Mezcua} et~al.}{2013b}]{2013MNRAS.436.2454M}
{Mezcua} M.,  {Lobanov} A.~P.,   {Mart{\'{\i}}-Vidal} I.,  2013b, \mn@doi
  [\mnras] {10.1093/mnras/stt1738}, \href
  {http://adsabs.harvard.edu/abs/2013MNRAS.436.2454M} {436, 2454}

\bibitem[\protect\citeauthoryear{{Mezcua}, {Roberts}, {Sutton}  \&
  {Lobanov}}{{Mezcua} et~al.}{2013c}]{2013MNRAS.436.3128M}
{Mezcua} M.,  {Roberts} T.~P.,  {Sutton} A.~D.,   {Lobanov} A.~P.,  2013c,
  \mn@doi [\mnras] {10.1093/mnras/stt1794}, \href
  {http://adsabs.harvard.edu/abs/2013MNRAS.436.3128M} {436, 3128}

\bibitem[\protect\citeauthoryear{{Mezcua}, {Fabbiano}, {Gladstone}, {Farrell}
  \& {Soria}}{{Mezcua} et~al.}{2014}]{2014ApJ...785..121M}
{Mezcua} M.,  {Fabbiano} G.,  {Gladstone} J.~C.,  {Farrell} S.~A.,   {Soria}
  R.,  2014, \mn@doi [\apj] {10.1088/0004-637X/785/2/121}, \href
  {http://adsabs.harvard.edu/abs/2014ApJ...785..121M} {785, 121}

\bibitem[\protect\citeauthoryear{{Mezcua}, {Roberts}, {Lobanov}  \&
  {Sutton}}{{Mezcua} et~al.}{2015}]{2015MNRAS.448.1893M}
{Mezcua} M.,  {Roberts} T.~P.,  {Lobanov} A.~P.,   {Sutton} A.~D.,  2015,
  \mn@doi [\mnras] {10.1093/mnras/stv143}, \href
  {http://adsabs.harvard.edu/abs/2015MNRAS.448.1893M} {448, 1893}

\bibitem[\protect\citeauthoryear{{Mezcua}, {Civano}, {Fabbiano}, {Miyaji}  \&
  {Marchesi}}{{Mezcua} et~al.}{2016}]{2016ApJ...817...20M}
{Mezcua} M.,  {Civano} F.,  {Fabbiano} G.,  {Miyaji} T.,   {Marchesi} S.,
  2016, \mn@doi [\apj] {10.3847/0004-637X/817/1/20}, \href
  {http://adsabs.harvard.edu/abs/2016ApJ...817...20M} {817, 20}

\bibitem[\protect\citeauthoryear{{Mezcua}, {Hlavacek-Larrondo}, {Lucey},
  {Hogan}, {Edge}  \& {McNamara}}{{Mezcua} et~al.}{2018a}]{2018MNRAS.474.1342M}
{Mezcua} M.,  {Hlavacek-Larrondo} J.,  {Lucey} J.~R.,  {Hogan} M.~T.,  {Edge}
  A.~C.,   {McNamara} B.~R.,  2018a, \mn@doi [\mnras] {10.1093/mnras/stx2812},
  \href {http://adsabs.harvard.edu/abs/2018MNRAS.474.1342M} {474, 1342}

\bibitem[\protect\citeauthoryear{{Mezcua}, {Civano}, {Marchesi}, {Suh},
  {Fabbiano}  \& {Volonteri}}{{Mezcua} et~al.}{2018b}]{2018MNRAS.478.2576M}
{Mezcua} M.,  {Civano} F.,  {Marchesi} S.,  {Suh} H.,  {Fabbiano} G.,
  {Volonteri} M.,  2018b, \mn@doi [\mnras] {10.1093/mnras/sty1163}, \href
  {http://adsabs.harvard.edu/abs/2018MNRAS.478.2576M} {478, 2576}

\bibitem[\protect\citeauthoryear{{Morganti}, {Tadhunter}, {Dickson}  \&
  {Shaw}}{{Morganti} et~al.}{1997}]{1997A&A...326..130M}
{Morganti} R.,  {Tadhunter} C.~N.,  {Dickson} R.,   {Shaw} M.,  1997, \aap,
  \href {http://adsabs.harvard.edu/abs/1997A%26A...326..130M} {326, 130}

\bibitem[\protect\citeauthoryear{{Morse}, {Cecil}, {Wilson}  \&
  {Tsvetanov}}{{Morse} et~al.}{1998}]{1998ApJ...505..159M}
{Morse} J.~A.,  {Cecil} G.,  {Wilson} A.~S.,   {Tsvetanov} Z.~I.,  1998,
  \mn@doi [\apj] {10.1086/306149}, \href
  {http://adsabs.harvard.edu/abs/1998ApJ...505..159M} {505, 159}

\bibitem[\protect\citeauthoryear{{Mortlock} et~al.,}{{Mortlock}
  et~al.}{2011}]{2011Natur.474..616M}
{Mortlock} D.~J.,  et~al., 2011, \mn@doi [\nat] {10.1038/nature10159}, \href
  {http://adsabs.harvard.edu/abs/2011Natur.474..616M} {474, 616}

\bibitem[\protect\citeauthoryear{{Nagar}, {Falcke}  \& {Wilson}}{{Nagar}
  et~al.}{2005}]{2005A&A...435..521N}
{Nagar} N.~M.,  {Falcke} H.,   {Wilson} A.~S.,  2005, \mn@doi [\aap]
  {10.1051/0004-6361:20042277}, \href
  {http://adsabs.harvard.edu/abs/2005A%26A...435..521N} {435, 521}

\bibitem[\protect\citeauthoryear{{Nguyen} et~al.,}{{Nguyen}
  et~al.}{2017}]{2017ApJ...836..237N}
{Nguyen} D.~D.,  et~al., 2017, \mn@doi [\apj] {10.3847/1538-4357/aa5cb4}, \href
  {http://adsabs.harvard.edu/abs/2017ApJ...836..237N} {836, 237}

\bibitem[\protect\citeauthoryear{{Nyland}, {Marvil}, {Wrobel}, {Young}  \&
  {Zauderer}}{{Nyland} et~al.}{2012}]{2012ApJ...753..103N}
{Nyland} K.,  {Marvil} J.,  {Wrobel} J.~M.,  {Young} L.~M.,   {Zauderer} B.~A.,
   2012, \mn@doi [\apj] {10.1088/0004-637X/753/2/103}, \href
  {http://adsabs.harvard.edu/abs/2012ApJ...753..103N} {753, 103}

\bibitem[\protect\citeauthoryear{{O'Dea}}{{O'Dea}}{1998}]{1998PASP..110..493O}
{O'Dea} C.~P.,  1998, \mn@doi [\pasp] {10.1086/316162}, \href
  {http://adsabs.harvard.edu/abs/1998PASP..110..493O} {110, 493}

\bibitem[\protect\citeauthoryear{{Pasham}, {Strohmayer}  \&
  {Mushotzky}}{{Pasham} et~al.}{2014}]{2014Natur.513...74P}
{Pasham} D.~R.,  {Strohmayer} T.~E.,   {Mushotzky} R.~F.,  2014, \mn@doi [\nat]
  {10.1038/nature13710}, \href
  {http://adsabs.harvard.edu/abs/2014Natur.513...74P} {513, 74}

\bibitem[\protect\citeauthoryear{{Penny} et~al.,}{{Penny}
  et~al.}{2018}]{2018MNRAS.476..979P}
{Penny} S.~J.,  et~al., 2018, \mn@doi [\mnras] {10.1093/mnras/sty202}, \href
  {http://adsabs.harvard.edu/abs/2018MNRAS.476..979P} {476, 979}

\bibitem[\protect\citeauthoryear{{Pezzulli}, {Valiante}  \&
  {Schneider}}{{Pezzulli} et~al.}{2016}]{2016MNRAS.458.3047P}
{Pezzulli} E.,  {Valiante} R.,   {Schneider} R.,  2016, \mn@doi [\mnras]
  {10.1093/mnras/stw505}, \href
  {http://adsabs.harvard.edu/abs/2016MNRAS.458.3047P} {458, 3047}

\bibitem[\protect\citeauthoryear{{Plotkin}, {Markoff}, {Kelly}, {K{\"o}rding}
  \& {Anderson}}{{Plotkin} et~al.}{2012}]{2012MNRAS.419..267P}
{Plotkin} R.~M.,  {Markoff} S.,  {Kelly} B.~C.,  {K{\"o}rding} E.,   {Anderson}
  S.~F.,  2012, \mn@doi [\mnras] {10.1111/j.1365-2966.2011.19689.x}, \href
  {http://adsabs.harvard.edu/abs/2012MNRAS.419..267P} {419, 267}

\bibitem[\protect\citeauthoryear{{Readhead}}{{Readhead}}{1994}]{1994ApJ...426...51R}
{Readhead} A.~C.~S.,  1994, \mn@doi [\apj] {10.1086/174038}, \href
  {http://adsabs.harvard.edu/abs/1994ApJ...426...51R} {426, 51}

\bibitem[\protect\citeauthoryear{{Reines}, {Greene}  \& {Geha}}{{Reines}
  et~al.}{2013}]{2013ApJ...775..116R}
{Reines} A.~E.,  {Greene} J.~E.,   {Geha} M.,  2013, \mn@doi [\apj]
  {10.1088/0004-637X/775/2/116}, \href
  {http://adsabs.harvard.edu/abs/2013ApJ...775..116R} {775, 116}

\bibitem[\protect\citeauthoryear{{Reines}, {Plotkin}, {Russell}, {Mezcua},
  {Condon}, {Sivakoff}  \& {Johnson}}{{Reines}
  et~al.}{2014}]{2014ApJ...787L..30R}
{Reines} A.~E.,  {Plotkin} R.~M.,  {Russell} T.~D.,  {Mezcua} M.,  {Condon}
  J.~J.,  {Sivakoff} G.~R.,   {Johnson} K.~E.,  2014, \mn@doi [\apjl]
  {10.1088/2041-8205/787/2/L30}, \href
  {http://adsabs.harvard.edu/abs/2014ApJ...787L..30R} {787, L30}

\bibitem[\protect\citeauthoryear{{Saikia}, {K{\"o}rding}, {Coppejans},
  {Falcke}, {Williams}, {Baldi}, {Mchardy}  \& {Beswick}}{{Saikia}
  et~al.}{2018}]{2018arXiv180506696S}
{Saikia} P.,  {K{\"o}rding} E.,  {Coppejans} D.~L.,  {Falcke} H.,  {Williams}
  D.,  {Baldi} R.~D.,  {Mchardy} I.,   {Beswick} R.,  2018, arXiv:1805.06696,
  \href {http://adsabs.harvard.edu/abs/2018arXiv180506696S} {}

\bibitem[\protect\citeauthoryear{{Secrest} et~al.,}{{Secrest}
  et~al.}{2015}]{2015ApJ...798...38S}
{Secrest} N.~J.,  et~al., 2015, \mn@doi [\apj] {10.1088/0004-637X/798/1/38},
  \href {http://adsabs.harvard.edu/abs/2015ApJ...798...38S} {798, 38}

\bibitem[\protect\citeauthoryear{{Secrest}, {Schmitt}, {Blecha}, {Rothberg}  \&
  {Fischer}}{{Secrest} et~al.}{2017}]{2017ApJ...836..183S}
{Secrest} N.~J.,  {Schmitt} H.~R.,  {Blecha} L.,  {Rothberg} B.,   {Fischer}
  J.,  2017, \mn@doi [\apj] {10.3847/1538-4357/836/2/183}, \href
  {http://adsabs.harvard.edu/abs/2017ApJ...836..183S} {836, 183}

\bibitem[\protect\citeauthoryear{{Smethurst} et~al.,}{{Smethurst}
  et~al.}{2016}]{2016MNRAS.463.2986S}
{Smethurst} R.~J.,  et~al., 2016, \mn@doi [\mnras] {10.1093/mnras/stw2204},
  \href {http://adsabs.harvard.edu/abs/2016MNRAS.463.2986S} {463, 2986}

\bibitem[\protect\citeauthoryear{{Soria}, {Hau}  \& {Pakull}}{{Soria}
  et~al.}{2013}]{2013ApJ...768L..22S}
{Soria} R.,  {Hau} G.~K.~T.,   {Pakull} M.~W.,  2013, \mn@doi [\apjl]
  {10.1088/2041-8205/768/1/L22}, \href
  {http://adsabs.harvard.edu/abs/2013ApJ...768L..22S} {768, L22}

\bibitem[\protect\citeauthoryear{{Stern} \& {Laor}}{{Stern} \&
  {Laor}}{2012}]{2012MNRAS.426.2703S}
{Stern} J.,  {Laor} A.,  2012, \mn@doi [\mnras]
  {10.1111/j.1365-2966.2012.21772.x}, \href
  {http://adsabs.harvard.edu/abs/2012MNRAS.426.2703S} {426, 2703}

\bibitem[\protect\citeauthoryear{{Sutton}, {Roberts}, {Walton}, {Gladstone}  \&
  {Scott}}{{Sutton} et~al.}{2012}]{2012MNRAS.423.1154S}
{Sutton} A.~D.,  {Roberts} T.~P.,  {Walton} D.~J.,  {Gladstone} J.~C.,
  {Scott} A.~E.,  2012, \mn@doi [\mnras] {10.1111/j.1365-2966.2012.20944.x},
  \href {http://adsabs.harvard.edu/abs/2012MNRAS.423.1154S} {423, 1154}

\bibitem[\protect\citeauthoryear{{Tadhunter}}{{Tadhunter}}{2016}]{2016A&ARv..24...10T}
{Tadhunter} C.,  2016, \mn@doi [\aapr] {10.1007/s00159-016-0094-x}, \href
  {http://adsabs.harvard.edu/abs/2016A%26ARv..24...10T} {24, 10}

\bibitem[\protect\citeauthoryear{{Terashima} \& {Wilson}}{{Terashima} \&
  {Wilson}}{2003}]{2003ApJ...583..145T}
{Terashima} Y.,  {Wilson} A.~S.,  2003, \mn@doi [\apj] {10.1086/345339}, \href
  {http://adsabs.harvard.edu/abs/2003ApJ...583..145T} {583, 145}

\bibitem[\protect\citeauthoryear{{Thean}, {Gillibrand}, {Pedlar}  \&
  {Kukula}}{{Thean} et~al.}{2001}]{2001MNRAS.327..369T}
{Thean} A.~H.~C.,  {Gillibrand} T.~I.,  {Pedlar} A.,   {Kukula} M.~J.,  2001,
  \mn@doi [\mnras] {10.1046/j.1365-8711.2001.04707.x}, \href
  {http://adsabs.harvard.edu/abs/2001MNRAS.327..369T} {327, 369}

\bibitem[\protect\citeauthoryear{{Ulvestad} \& {Ho}}{{Ulvestad} \&
  {Ho}}{2001}]{2001ApJ...558..561U}
{Ulvestad} J.~S.,  {Ho} L.~C.,  2001, \mn@doi [\apj] {10.1086/322307}, \href
  {http://adsabs.harvard.edu/abs/2001ApJ...558..561U} {558, 561}

\bibitem[\protect\citeauthoryear{{Volonteri} \& {Rees}}{{Volonteri} \&
  {Rees}}{2005}]{2005ApJ...633..624V}
{Volonteri} M.,  {Rees} M.~J.,  2005, \mn@doi [\apj] {10.1086/466521}, \href
  {http://adsabs.harvard.edu/abs/2005ApJ...633..624V} {633, 624}

\bibitem[\protect\citeauthoryear{{Volonteri}, {Haardt}  \& {Madau}}{{Volonteri}
  et~al.}{2003}]{2003ApJ...582..559V}
{Volonteri} M.,  {Haardt} F.,   {Madau} P.,  2003, \mn@doi [\apj]
  {10.1086/344675}, \href {http://adsabs.harvard.edu/abs/2003ApJ...582..559V}
  {582, 559}

\bibitem[\protect\citeauthoryear{{Webb} et~al.,}{{Webb}
  et~al.}{2012}]{2012Sci...337..554W}
{Webb} N.,  et~al., 2012, \mn@doi [Science] {10.1126/science.1222779}, \href
  {http://adsabs.harvard.edu/abs/2012Sci...337..554W} {337, 554}

\bibitem[\protect\citeauthoryear{{Wilson} \& {Tsvetanov}}{{Wilson} \&
  {Tsvetanov}}{1994}]{1994AJ....107.1227W}
{Wilson} A.~S.,  {Tsvetanov} Z.~I.,  1994, \mn@doi [\aj] {10.1086/116935},
  \href {http://adsabs.harvard.edu/abs/1994AJ....107.1227W} {107, 1227}

\bibitem[\protect\citeauthoryear{{Wrobel} \& {Ho}}{{Wrobel} \&
  {Ho}}{2006}]{2006ApJ...646L..95W}
{Wrobel} J.~M.,  {Ho} L.~C.,  2006, \mn@doi [\apjl] {10.1086/507102}, \href
  {http://adsabs.harvard.edu/abs/2006ApJ...646L..95W} {646, L95}

\bibitem[\protect\citeauthoryear{{Wu} et~al.,}{{Wu}
  et~al.}{2015}]{2015Natur.518..512W}
{Wu} X.-B.,  et~al., 2015, \mn@doi [\nat] {10.1038/nature14241}, \href
  {http://adsabs.harvard.edu/abs/2015Natur.518..512W} {518, 512}

\bibitem[\protect\citeauthoryear{{Yang} et~al.,}{{Yang}
  et~al.}{2017}]{2017MNRAS.464L..70Y}
{Yang} X.,  et~al., 2017, \mn@doi [\mnras] {10.1093/mnrasl/slw160}, \href
  {http://adsabs.harvard.edu/abs/2017MNRAS.464L..70Y} {464, L70}

\bibitem[\protect\citeauthoryear{{van Wassenhove}, {Volonteri}, {Walker}  \&
  {Gair}}{{van Wassenhove} et~al.}{2010}]{2010MNRAS.408.1139V}
{van Wassenhove} S.,  {Volonteri} M.,  {Walker} M.~G.,   {Gair} J.~R.,  2010,
  \mn@doi [\mnras] {10.1111/j.1365-2966.2010.17189.x}, \href
  {http://adsabs.harvard.edu/abs/2010MNRAS.408.1139V} {408, 1139}

\makeatother
\end{thebibliography}

%%%%%%%%%%%%%%%%%%%%%%%%%%%%%%%%%%%%%%%%%%%%%%%%%%
%%%%%%%%%%%%%%%%%%%%%%%%%%%%%%%%%%%%%%%%%%%%%%%%%%

% Don't change these lines
\bsp	% typesetting comment
\label{lastpage}
\end{document}